\documentclass[11pt]{ieeeconf} 
\usepackage{cite}
\usepackage{amsmath,amssymb,amsfonts}
\usepackage{algorithm,algpseudocode}
\usepackage{graphicx}
\usepackage{textcomp}
\usepackage{times,subfigure,url,longtable,ifthen,subcaption,balance}
\newboolean{showcomments}
\setboolean{showcomments}{true}
\newcommand{\Modified}[1]{\ifthenelse{\boolean{showcomments}} {\textcolor{red}{#1}}{}}
\usepackage{array}
\newcolumntype{M}[1]{>{\centering\arraybackslash}m{#1}}
\newcolumntype{P}[1]{>{\centering\arraybackslash}p{#1}}

\title{Empowering the Grid: Decentralized Autonomous Control for Effective Utilization and Resilience
\thanks{This work is currently under review in a Journal. \\ 
This research was supported by the Resilience through Data-driven Intelligently-Designed Control (RD2C) Initiative at Pacific Northwest National Laboratory (PNNL). PNNL is operated for the U.S. Department of Energy (DOE) by Battelle Memorial Institute under Contract No. DE-AC05-76RL01830. 
Emails:
  \{saipushpak.n,ak.bharati,subhrajit.sinha, soumya.kundu,veronica.adetola,kevin.schneider\}@pnnl.gov} 
}

\author{
Sai~Pushpak~Nandanoori, Alok~Kumar~Bharati,
Subhrajit~Sinha, Soumya~Kundu, Veronica~Adetola,  Kevin~Schneider \\ 
  Pacific Northwest National Laboratory \\ 
  Richland, WA 99354 USA 
}

\begin{document}
\maketitle

\begin{abstract}
With the emergence of low-inertia microgrids powered by inverter-based generation, there remains a concern about the operational resilience of these systems. Grid-forming inverters (GFMs), enabled by various device-level (primary) and system-level (secondary) control methods, are poised to play a significant role in achieving certain operational objectives, such as the effective utilization of clean energy resources while maintaining stability. However, despite the recent advances in GFMs, there is a lack of suitable controls that can ascertain resilience-constrained operations, like maintaining critical operational safety limits during transients under various cyber-physical disruptions. In this work, we develop decentralized autonomous controllers (DACs) that enforce resilience-constrained operation via local, minimally invasive adjustments (e.g., changes in set-points) while co-existing within the hierarchy of existing (primary and secondary) controls. The DACs work autonomously by sensing only local GFM measurements and act only when operational resilience constraints are violated. The proposed DAC scheme is computationally efficient (only algebraic computations), which enables fast, real-time execution and demonstrates the efficacy of the proposed control framework on GridLAB-D-HELICS-based control-grid co-simulations on the IEEE 123-node networked microgrid. Finally, we show how the developed DACs empower the grid by utilizing the available resources entirely to ensure resilience (maintain frequency safe limits). 
\end{abstract}

\section{Introduction}

Microgrids 
are gaining significant traction as a way to improve energy security and resilience and integrate renewable energy sources. These localized grids can operate independently (islanded mode) or connect to the main grid. At the heart of renewable-rich microgrids are inverter-based resources, playing a critical role in maintaining stability. Moreover, the distribution systems and, in particular, the microgrids are becoming more and more complex with the rapid integration of distributed energy resources, controls, communication schemes, regulatory policies, etc. \cite{schneider2020networked}.

Furthermore, as the cyber-physical power systems grow in complexity and size, with often different parts of the network being operated independently by different stakeholders \cite{norouzi2022review}, there is a strong need for advanced algorithms that allow multiple decision-making agents within the network to act autonomously to enforce resilience under different cyber-physical adversarial events. 
 
GFMs have emerged as a key component {for} effectively utilizing clean and renewable energy resources in microgrids while providing stability (e.g., damping) \cite{lasseter2019grid}. In the hierarchy of controls for GFM inverters \cite{palizban2015hierarchical}, the power-frequency droop control sits at the primary layer, adjusting the inverter (real/active) power injected into the grid to stabilize (dampen) the frequency oscillations during any disturbance. It is important to notice that the stabilized frequency needs to be at the nominal frequency, which is ensured by the secondary layer \cite{palizban2015hierarchical} operating at slower timescales and adjusting the real power reference (set-points) to the inverters across the grid.  

Inverter-based resources bring forth novel challenges for the hierarchical layer of controls used to operate today's power grid \cite{gu2022power}. One of these challenges is manifested in fast frequency dynamics, driven by low inertia and fast actions of inverter controls. As a result, there is an increasing overlap of the timescales of operation, with implications on the validity of the assumption around decoupled control layers \cite{taylor2016power}, as well as on various frequency-based protections (e.g., under-frequency thresholds, rate-of-change-of-frequency) \cite{gu2022power,muenz2024protection}. Existing primary (local, device-level) and secondary (centralized/distributed, system-level) controls within the cyber-physical power systems today lack the constructive design methods to enforce operational safety limits (e.g., under- and over-frequency limits as per IEEE Standard 1547 \cite{ieee1547}) that ensure operational resilience during transients. 

Control of inverters, including GFMs, has attracted immense attention in recent years within the power and control communities. Recent research into primary (local, device-level) controls of inverters has led to novel control designs \cite{jiang2020dynamic,geng2022unified}, as well as methods for tuning controller gains for stability \cite{kundu2019identifying,nandanoori2020distributed}. However, the primary controls are designed and operated myopically, without any mechanism to actively enforce resilience via local measurement and control action (see \cite{guerrero2012advanced,rathnayake2021grid} and the references therein). 

On the other hand, the secondary (centralized/distributed, system-level) control helps restore the grid's frequency to the nominal level, using various methods based on optimal control \cite{khayat2018decentralized,lian2023decentralized}, consensus protocols \cite{singhal2022consensus}, passitivity \cite{liu2023decentralized}, and blockchain-enabled controls with neural networks \cite{veerasamy2023blockchain}. However, the secondary controls operate at a slower timescale and are incapable of ensuring operational resilience during transients, especially in the presence of frequent stochastic disturbances or unforeseen adversarial events. Moreover, many adversarial events in critical cyber-physical networks can propagate as a cascading failure, which often starts from local disruptions that travel through the network over a short period and can create a system-wide impact (e.g., a blackout) \cite{Madhu_presentation}. 

Some of the drawbacks of these methods include their dependency on a specific form of the inverter model and the control design is based on linearization, which significantly impacts the applicability of the controls under various operating conditions and does not enforce resilience constraints into the control design to achieve resilience irrespective of the nature of the disturbance (see \cite{weitenberg2018robust} for a comprehensive study of merits and trade-offs of decentralized secondary controls). Moreover, most of the controls develop different secondary controls to achieve nominal frequency and power sharing among the inverters, which also becomes a challenge as it will be difficult to finalize a secondary control strategy for real-world application. 

These challenges further necessitate the need to develop agile, adaptive, and lightweight control solutions that allow the control agents to autonomously (under no communication) ensure system-wide resilience under various disturbances. Such solutions, as proposed in this work, should be
\begin{enumerate}
    \item Minimally invasive, that is, minimal changes to the existing control architecture, so it is easy to implement and integrate with the existing control hierarchy,  
    \item Decentralized and autonomous (few measurements required to process controls), 
    \item Computationally efficient (ability to react fast), and 
    \item Largely model-agnostic, i.e., not tied to any specifics of the primary/device-level controls. 
\end{enumerate}

In this work, we develop a solution similar to the one introduced in \cite{kundu2020transient} to intervene only when certain local resilience constraints are not met, and this control strategy works in tandem along an existing primary and secondary control hierarchy. We first begin by defining a safe operating region for frequency and resilience as the ability of the system to maintain this safe region irrespective of the nature of the disturbance. These controls are referred to as safety-promoting DACs, and these DACs were designed offline based on inverter nameplate details, droop gains, etc., and implemented in real-time where the inverter measurements such as the power and frequency measurements are continuously monitored and the set-points are adjusted accordingly. The DAC does not act during small disturbances and comes into play only if the frequency deviates outside the safe region. We achieve this by defining a barrier function that comes into play only when the frequencies are violating the safe limits. This work was inspired by the knowledge gained from previous works \cite{kundu2019distributed,kundu2020transient} that involved solving multiple SOS-optimization problems (which we overcome in this work and make it computationally efficient) to provide sufficient guarantees on maintaining safe region.  

The main contributions of this work are described as follows:
\begin{enumerate}
    \item Defining barrier functions for each GFM as a function of the local operational resilience constraints. 
    \item Developing a decentralized set-point change control strategy by applying barrier functions to local measurements to address operational resilience violations. 
    \item Providing sufficient guarantees on meeting the operational constraints while being agnostic to the nature of the disturbance.
    \item Demonstration of the proposed decentralized control functionality under critical operating conditions in a networked microgrid system.
\end{enumerate}

A key observation was the fact that the addition of the DACs results in the enhancement of the current functionality of GFMs by empowering these vital resources to act with urgency when needed. Therefore, it is important to notice that these controls complement the existing hierarchy of controls and act only when local operational resilience conditions such as frequency or voltage limits are violated. 

The rest of the paper is organized as follows. Section~\ref{sec:GFM_model} describes the existing hierarchy of controls and shows where the proposed DACs fit in. The design of DACs is presented in Section~\ref{sec:DAC}, and an application of DACs to 123-node control-grid-co-simulations follows in Section~\ref{sec:sim_study}. Finally, the findings and conclusions of this work are discussed in Section~\ref{sec:Con}.

\section{Grid Forming Inverter Model and Inverter Controls}
\label{sec:GFM_model}
\subsection{Grid Forming Inverter}
\begin{figure}[ht]
    \centering
    \includegraphics[width=0.95\columnwidth]{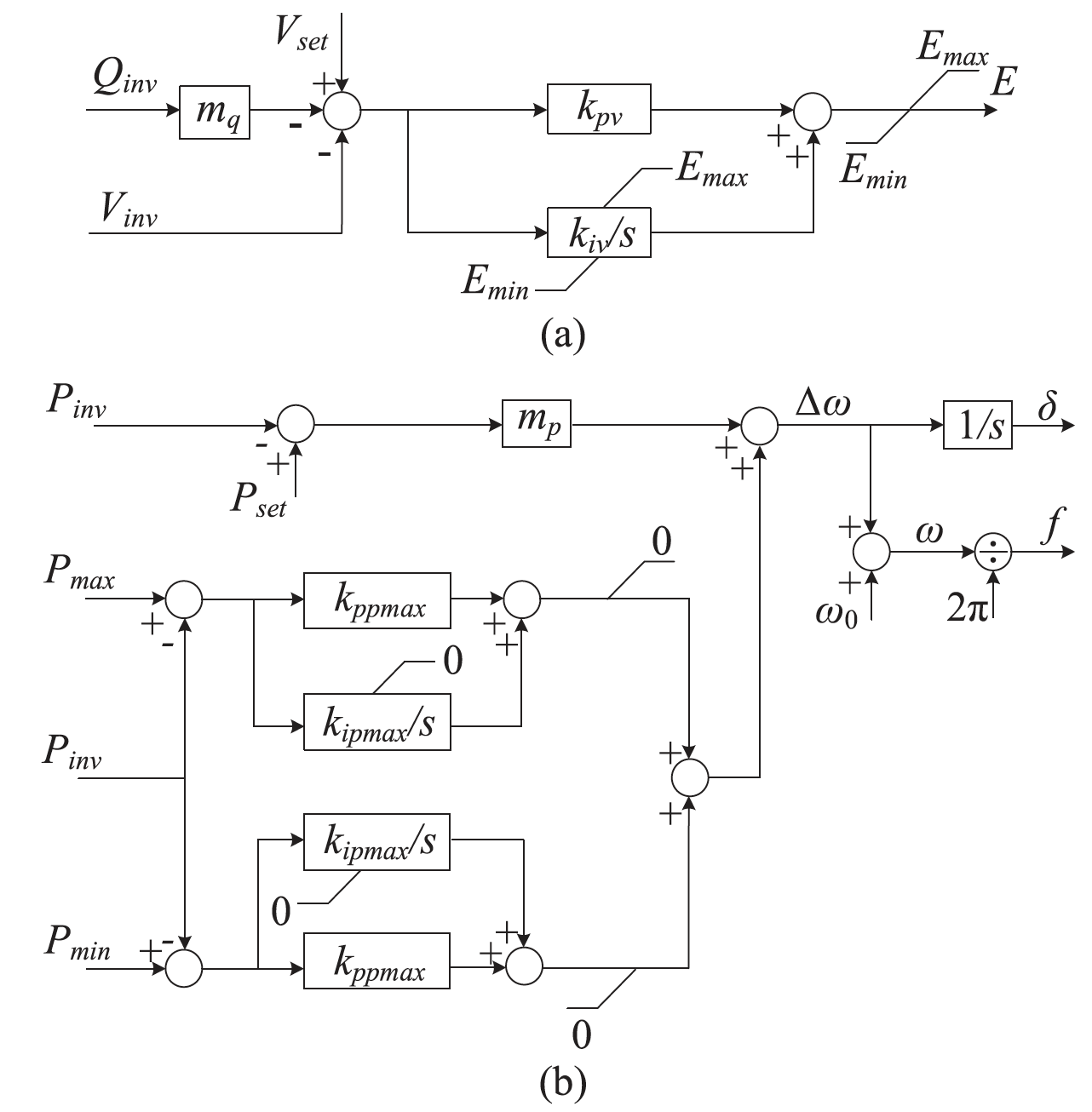}
    \caption{CERTS droop-controlled GFM inverter model, with (a) Q-V droop control, and (b) P-f droop control, including over- and under-load mitigation (adopted from \cite{du2020modeling}). }
    \label{fig:GFM_PI}
\end{figure}

Consider the GFM inverter with CERTS droop-controlled GFM model from \cite{lasseter2010certs,du2020modeling} as shown in Fig. \ref{fig:GFM_PI}. The GFM inverter model consists of a P-f droop control, Q-V droop control as well as over- and under-load mitigation strategy. 

In addition, there are low-pass measurement filters that measure the active (P), reactive (Q) power, and voltage at the point of common coupling for the inverter. $P_{\textrm{grid},\phi}$ and $Q_{\textrm{grid},\phi}$ are the active and reactive power injections from the inverter into the grid, at phase $\phi\in\lbrace a,b,c\rbrace$\, while $V_{\textrm{grid},\phi}$ is the voltage at phase $\phi\in\lbrace a,b,c\rbrace$\,. 
The inverters also honor power rating constraints; therefore, 
\begin{align*}
    P_{\textrm{grid}}\in\left[P_\textrm{min},P_\textrm{max}\right]
\end{align*}
With the assumption that $\dot{P}_{\textrm{set}}=0$ (i.e., the $P_{\textrm{set}}$ is changed {slowly} compared to the inverter droop control timescales), we have the P-f droop equations given by
\begin{align}
     \dot{\delta}&=\omega - \omega_0\\
\tau \dot{\omega} &= -\left(\omega-\omega_0\right) + m_p\left(P_{\textrm{set}}-P_{\textrm{inv}}\right) \label{eq:pf_droop}
\end{align}    
\noindent where $P_{\textrm{inv}} := {\sum}_{\phi\in\lbrace a,b,c\rbrace}P_{\textrm{grid},\phi}$, $\omega_0$ is the nominal frequency, $m_p$ is the P-f droop gain and $P_{\textrm{set}}$ is the input to the inverter dynamics. As mentioned earlier, the $P_{set}$ dynamics and the inverter droop control dynamics operate at different timescales, and as such there are separate control strategies for controlling the inverter dynamics (primary control) and $P_{set}$ dynamics (secondary control) \cite{guerrero2010hierarchical}.

\subsubsection*{Primary Controls}
The P-f droop control at the GFM inverters acts as a primary control to the inverters and operates at millisecond ($ms$) timescales. The P-f droop controls (based on the internal inverter measurements) react quickly to any disturbance in the system to stabilize the frequency.  The input to the droop controls is the reference set-point ($P_\textrm{set}$ or $Q_\textrm{set}$), which is usually provided by the secondary controls. The action of the droop controls in most cases will stabilize the frequency; however, there are no guarantees that the frequency will go back to the nominal frequency ($60$ Hz). 

\subsubsection*{Secondary Controls} The real power set-points to the inverters are updated by the secondary controls by accessing the inverter measurements. These controls could be either centralized or distributed and operate at slower timescales (in the order of a few seconds), compared to primary controls. The objective of this control is to change the real power set-points in such a way that the frequency reaches nominal with equal power sharing among the inverters. {The secondary controls discussed in \cite{singhal2022consensus} are implemented in this work.} These controls are based on the leader-follower consensus framework and coordinate the GFM and grid-following (GFL) inverters to achieve frequency restoration to nominal frequency and accurate power sharing. 

\begin{figure*}[ht!]
    \centering
    \includegraphics[width = 0.95 \textwidth]{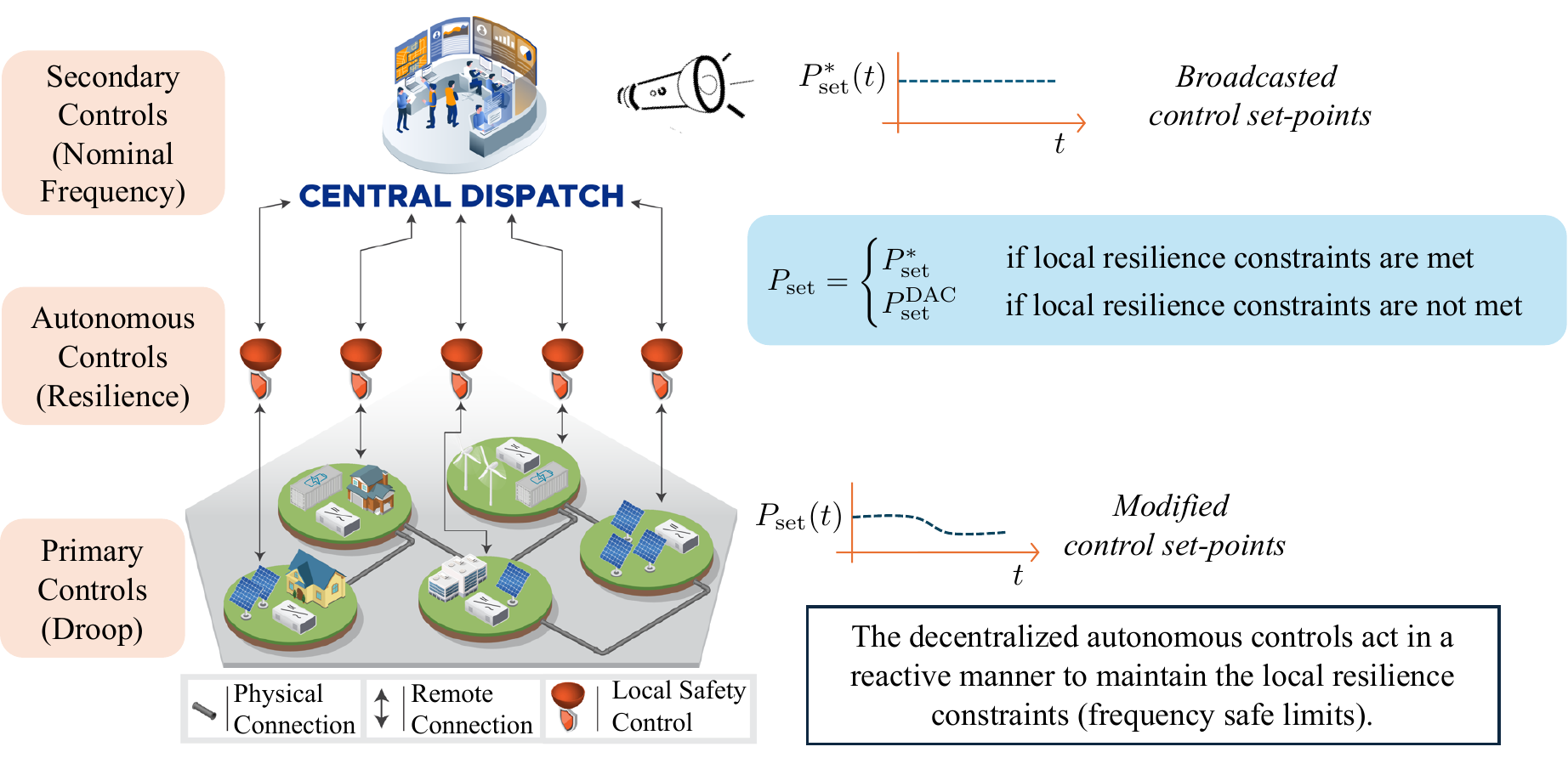}    
    \caption{Overview of the functioning of the proposed DAC: These inverter-based controls are located between primary and secondary controls. When the frequency resilience constraints are violated, these controls adjust the secondary control set-points to ensure frequency resilience, otherwise, these controls {will not} intervene and pass the secondary control set-points to the inverter. These autonomous controls are computationally efficient and rely solely on local measurements available at the inverters themselves.}
    \label{fig:control_hierarchy}
\end{figure*}

In the scope of this work, we define a safe operating region around the nominal frequency and define resilience as the ability of the system to maintain this safe region irrespective of the type of events happening on the system. Considering the timescale gap between the primary controls and secondary controls of operation there is a blind spot and if an event happens during this period the primary controls react but cannot ensure frequency resilience that could avoid any violations that lead to blackouts. To bridge the timescale gap between the primary and secondary controls and achieve resilience, we introduce reactive controls that are decentralized and maintain resilience.

The functioning of the DACs is illustrated in Fig.~\ref{fig:control_hierarchy} and is described in detail in the following section. 

\section{Decentralized Autonomous Controls}
\label{sec:DAC}
In this section, we present a decentralized autonomous control framework that utilizes a few local measurements from the inverters to maintain a predefined safe region for frequency under disturbances. We begin by defining the local resilience constraints at each inverter. 

At each inverter, suppose the frequency limits are defined as $\omega \in [\omega_{\textrm{min}}, \omega_{\textrm{max}}]$ where $\omega_{\textrm{min}}, \omega_{\textrm{max}}$ are the lower and upper-frequency limits respectively. These pre-defined safe limits for frequency at each inverter form the local resilience constraints that need to be maintained irrespective of the nature of the disturbance. Therefore, the goal is to modify the set points $P_{\textrm{set}}$, such that the local resilience constraints are respected. To this end, we introduce two barrier functions, corresponding to lower and upper frequency limits, that act when frequencies deviate from the safe limits either from below or above. 

Consider the following barrier functions defined as, 
\begin{align}
    B_{\textrm{min}}(\omega) = \omega - \omega_{\textrm{min}}, \, \mbox{such that}\, \begin{cases}
        \geq 0, & \mbox{Safe} \\ 
        < 0, & \mbox{Unsafe}
    \end{cases} \label{eq:b_min} \\
    B_{\textrm{max}}(\omega) = \omega - \omega_\textrm{max}, \, \mbox{such that}\, \begin{cases}
        \leq 0, & \mbox{Safe} \\ 
        > 0, & \mbox{Unsafe}
    \end{cases} \label{eq:b_max}
\end{align}
We refer to $B_{\textrm{min}}$ and $B_{\textrm{max}}$ as the lower and upper barrier functions respectively. When the lower (upper) barrier function is positive (negative), the lower (upper) frequency safe limit is not violated. However, in the presence of a disturbance, if the lower frequency limit is violated, we want the derivative of the lower barrier function to change its sign and become non-decreasing. Similarly, due to a disturbance, if the upper frequency limit is violated, we want the derivative of the upper barrier function to change its sign and become non-increasing. Thus we need the following conditions on the time-derivative of the barrier functions. 
\begin{align}
    \dot{B}_{\textrm{min}} \geq -\alpha B_{\textrm{min}}^q, 
    \label{eq:dBmin_dt} \\ 
    \dot{B}_{\textrm{max}} \leq -\alpha B_{\textrm{max}}^q, 
    \label{eq:dBmax_dt} 
\end{align}
where $q$ is required to be an odd number to preserve the sign of the right hand side of Eqs.~\eqref{eq:dBmin_dt}-\eqref{eq:dBmax_dt} and $\alpha$ can be considered as control performance gain. The addition of the term, $-\alpha B_{\textrm{min}}^q$ ($-\alpha B_{\textrm{max}}^q$) on the right side of the barrier function derivative requires explanation. Whenever the frequencies are inside the safe region, the lower barrier function is nonnegative and we can allow the lower barrier function to decrease if the operating frequency is well inside the safe region. However, when the operating frequency goes outside the safe region from below, we want the lower barrier function to increase and the corresponding time-derivative to be non-negative (that is, $\dot{B}_{\textrm{min}} \geq 0$). 

Similarly, the non-positive values of the upper barrier function indicate the frequency is within the safety limits, and the upper barrier function {can} increase inside the safe region as long as the operating frequency is lower than $\omega_{\textrm{max}}$. However, when the operating frequency increases beyond, $\omega_{\textrm{max}}$, the upper barrier function should decrease and the corresponding time-derivative should be non-positive (that is, $\dot{B}_{\textrm{max}} \leq 0$).

When the frequencies deviate outside the safe region due to an event, it is expected to steer the system to the safe region in finite time and to achieve this, we modify, $\dot{B}_{\textrm{min}} \geq 0$ to $\dot{B}_{\textrm{min}} \geq -\alpha B_{\textrm{min}}^q$ and similarly, $\dot{B}_{\textrm{max}} \leq 0$ to $\dot{B}_{\textrm{max}} \leq -\alpha B_{\textrm{max}}^q$. To achieve the desired performance to maintain safety while ensuring the smoothness of the resulting control adjustment, we chose $q=3$. 
Therefore, from Eq.~\eqref{eq:dBmin_dt}, we have, 
\begin{align}
    & P_{\textrm{set}} \geq P_{\textrm{inv}} + \frac{1}{m_p} \left(\omega-\omega_0 -\alpha (\omega-\omega_{\textrm{min}})^3\right) \label{eq:P_set_low} \\
    & \implies \omega \geq \omega_{\textrm{min}} 
 \quad (\mbox{lower limit}) \nonumber
\end{align}
Similarly, from Eq.~\eqref{eq:dBmax_dt}, we have, 
\begin{align}
    & P_{\textrm{set}} \leq P_{\textrm{inv}} + \frac{1}{m_p} \left(\omega-\omega_0 -\alpha (\omega-\omega_{\textrm{max}})^3\right) \label{eq:P_set_up} \\
    & \implies \omega \leq \omega_{\textrm{max}} 
 \quad (\mbox{upper limit}) \nonumber
\end{align}
The set-point changes suggested by Eq.~\eqref{eq:P_set_low} and Eq.~\eqref{eq:P_set_up} result in a frequency that is within the safe region. {However, as the scope of this work is only concerned with maintaining the safe region and allowing the secondary control to drive the frequency to the nominal value,} it is sufficient to make minimal set-point changes such that either $\omega = \omega_{\textrm{min}}$ or $\omega = \omega_{\textrm{max}}$ is maintained. This results in the following conditions. 
\begin{equation}
\begin{aligned}
    P_{\textrm{set}}^{\textrm{low}} = & P_{\textrm{inv}} + \frac{1}{m_p} \left(\omega-\omega_0 -\alpha (\omega-\omega_{\textrm{min}})^3\right) \\ 
    & \implies \omega = \omega_{\textrm{min}} 
\end{aligned}
\label{eq:P_set_low_eq}
\end{equation}
\begin{equation}
\begin{aligned}
    P_{\textrm{set}}^{\textrm{up}}  = & P_{\textrm{inv}} + \frac{1}{m_p} \left(\omega-\omega_0 -\alpha (\omega-\omega_{\textrm{max}})^3\right)\\ 
    & \implies \omega = \omega_{\textrm{max}} 
\end{aligned}
\label{eq:P_set_up_eq}
\end{equation}
When any event on the system drives the frequencies outside the safe region either from below (under frequency event) or above (over frequency event), the set-points to the GFM inverters are adjusted accordingly either by Eq.~\eqref{eq:P_set_low_eq} or Eq.~\eqref{eq:P_set_up_eq} to maintain the safe region. 

\begin{figure}[h!]
    \centering
    \includegraphics[width = 0.95 \columnwidth]{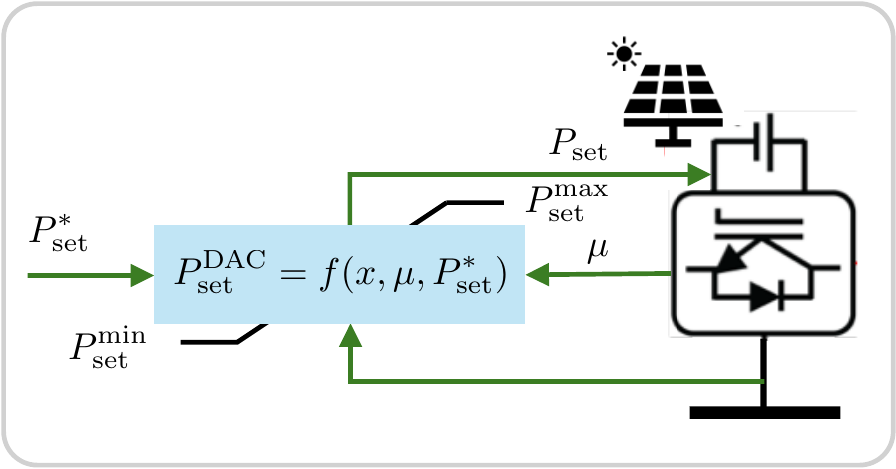}
    \caption{DAC implementation at each inverter. $P_\textrm{set}^*$ denotes the set-point sent by the secondary control, $\mu = (S_{\textrm{inv}}, m_p)$ denotes the fixed parameters at the inverters and $x = (\omega, P_{\textrm{inv}}, Q_{\textrm{inv}})$ denote the measurements from the inverter.}
    \label{fig:inv_control_implementation}
\end{figure}

{These two cases are combined to deploy these safety controls at the inverters.} Let $P_\textrm{set}^*$ be the set point given by the secondary controls or central dispatch to the inverter and let $P_{\textrm{set}}$ be the set-point seen by the inverter. Since the safe controls always act to maintain the safe region, we have, 
\begin{align}
    P_\textrm{set} = \begin{cases}
        P_{\textrm{set}}^* & \; \mbox{if }\; \omega_{\textrm{min}} \leq \omega \leq \omega_{\textrm{max}} \\
        P_{\textrm{set}}^{\textrm{DAC}} & \; \mbox{otherwise}
     \end{cases}
\end{align}
where 
\begin{align}
    P_{\textrm{set}}^{\textrm{DAC}} = min(P_{\textrm{set}}^{\textrm{up}}, max(P_{\textrm{set}}^{\textrm{low}}, P_{\textrm{set}}^*))    \label{eq:P_set_DAC}
\end{align}
where $P_{\textrm{set}}^{\textrm{DAC}}$ denotes the set-points that are modified by the safety-promoting DACs. Note that the control performance gain $\alpha$ is chosen in such a way that these safety-promoting DACs won't act if frequency violations will not happen. Therefore, when $\omega_\textrm{min} \leq \omega \leq \omega_\textrm{max}$, $P_\textrm{set} = P_\textrm{set}^*$, which essentially indicates that the real power set-point provided by secondary controls will be passed to the inverter. {Furthermore, the following criterion ensures that} the DACs set-point changes respect the capacity constraints, we have the following criterion. 
\begin{align}
    P_\textrm{\textrm{set}}^{\textrm{max}} = & \sqrt{S_{\textrm{inv}}^2 - Q_{\textrm{inv}}^2} \nonumber \\
    P_\textrm{set} = & min(P_\textrm{set}^{\textrm{max}},max(P_\textrm{set}^{\textrm{min}},P_\textrm{set})), 
\end{align}
where $Q_{\textrm{inv}} = {\sum}_{\phi\in\lbrace a,b,c\rbrace}Q_{\textrm{grid},\phi}$, and $S_{\textrm{inv}}$ is the size of the inverter. The implementation of the safety-promoting DACs is illustrated in Fig. \ref{fig:inv_control_implementation}. The DACs use only a few measurements from the inverters to adjust the real power set-points at the inverters and maintain the safe frequency region (in other words, local resilience). The real power control set-points are modified according to Eq.~\eqref{eq:P_set_DAC} which is an algebraic equation, and as a consequence, these controls are computationally efficient. Therefore, due to the decentralized autonomous and computationally efficient properties of the DACs, they are more suitable for real-world applications. 

\begin{figure}[h!]
    \centering
    \includegraphics[width = 0.95 \columnwidth]{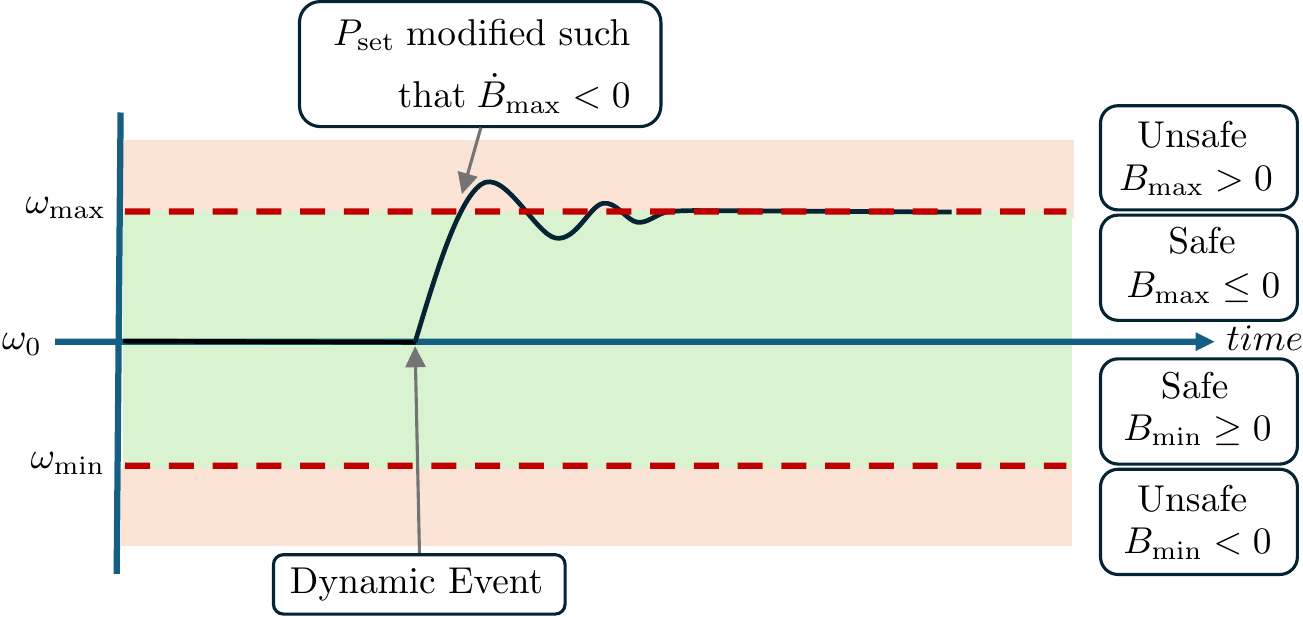}
    \caption{Illustration of the functioning of safety promoting DACs when frequency violations occur. }
    \label{fig:DACs_functionality_overview}
\end{figure}

Fig. \ref{fig:DACs_functionality_overview} provides an overview of how the DACs function during the onset of a dynamic event. When a dynamic event occurs that results in an over-frequency event and the safe region is violated, the control set-points are modified in such a way that the frequency safe region is maintained (that is, $P_\textrm{set}^{\textrm{DAC}}$ modified such that $\dot{B}_\textrm{max} \leq 0$). 

The following algorithmic procedure encapsulates the entire DAC implementation methodology applicable to any GFM inverter.

\begin{algorithm}[hbt!]
\caption{DAC algorithm to compute the new set-point at any GFM inverter}\label{alg:DAC}
\begin{algorithmic}
\Require Fixed parameters $S_{\textrm{inv}}, m_p, \alpha$
\State \hspace{-0.35 cm}\textbf{Inputs:} $\omega, P_{\textrm{inv}}, Q_{\textrm{inv}}, P_{\textrm{set}}^*$
\State Compute $P_{\textrm{set}}^{\textrm{low}}, P_{\textrm{set}}^{\textrm{up}}$ from Eqs.~(9)-(10)
\State Find $P_{\textrm{set}}$ from Eq.~(11). 
\State Use Eq.~(13) to ensure $P_{\textrm{set}}$ is within the absolute limits.
\State \hspace{-0.4 cm} \textbf{Outputs:} $P_{\textrm{set}}$
\end{algorithmic}
\end{algorithm}

\subsection{Control Performance Design Guidelines}
The control performance parameters, $\alpha$ and $q$ are introduced to ensure certain conditions on the time-derivative of the barrier functions which results in appropriate control setpoint changes to guarantee the safe frequency region. Since we want the control to act minimally inside the safe bounds and act significantly when outside the safe region, we cannot have a constant rate of change for the derivatives of $B_{min}$ and $B_{max}$, and hence the derivatives of $B_{min}$ and $B_{max}$ need to be nonlinear as a function of the frequency deviation. This is achieved by introducing $q$ and theoretically, $q$ can be any odd natural number.


However,  for large $q$, $B_{min}^q$ or $B_{max}^q$ grows exponentially fast for large deviations in frequency and makes the control extremely aggressive. Thus, for practical consideration, we cannot choose large $q$. Therefore, for large deviations in $\omega$, $\dot{B}_{min}$ and $\dot{B}_{max}$ becomes extremely large. On the other end, for $q=1$, the derivatives of the barrier function remain constant both inside and outside the safe region. Hence, we chose $q=3$, the smallest odd natural number greater than one. 

To control the time-derivative of barrier functions, the control performance gain, $\alpha>0$ is introduced in Eqs.~\eqref{eq:dBmin_dt}-\eqref{eq:dBmax_dt}. For a given $\alpha > 0$, $B_{min}$ or $B_{max}$ reaches steady state exponentially. In practical scenarios, the value $\alpha$ is adjusted to magnify the frequency error, thereby enabling smoother control. Consequently, the selection of $\alpha$ depends on the chosen value of $q$. For instance, consider $q=3$, $\omega_{\textrm{min}} = 59.5 Hz$, $\omega_{\textrm{max}} = 60.5 Hz$ and current $\omega = 59.4 Hz$. Then $\alpha$ must be at least $\vert (59.4-59.5)^3\vert = 0.001$. If $\alpha \leq 0.001$, then the control does not see any deviation from the safe region and hence won't act. However, suppose if $\alpha >0.001$, then the controls would start acting and the control action becomes aggressive if $\alpha$ is larger. 

The following section demonstrates the use of DACs to empower the grid operation and ensure resilience under various operating conditions.

\section{Demonstration of the DAC Functionality Under Various System Operating Conditions}
\label{sec:sim_study}
\begin{figure}[h!]
    \centering
    \includegraphics[width = 0.95 \columnwidth]{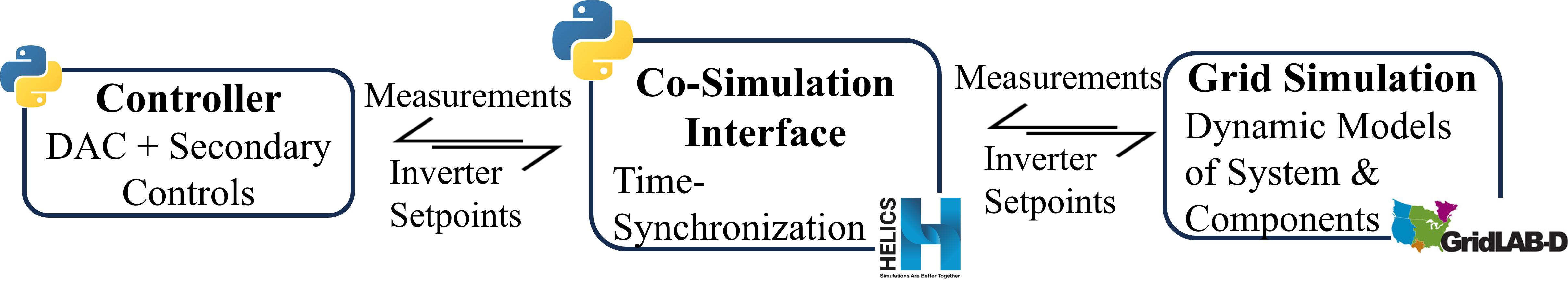}
    \caption{Control-grid co-simulation framework used for the simulation study.}
    \label{fig:cosim}
\end{figure}

This section details how the DACs are integrated with the grid simulations through the control-grid co-simulations. Fig. \ref{fig:cosim} shows the main components of the control-grid co-simulation framework (for details we refer the reader to \cite{bharati2021scalable,bharati2021smtd}). 
In the following, we discuss the various operating conditions of the system considered and present the simulation results while demonstrating the versatility of the proposed DAC. 

\subsection{IEEE 123-node System Model}
The authors in \cite{schneider2023adaptive} considered the standard IEEE 123-node test system and modified it to form four independent regions of which 3 can be formed as microgrids using diesel generators (DGs) and GFLs. The redesigned system features three independent, self-regulating microgrids alongside a PV-rich zone reliant on external support for stable operation. This results in three microgrids and a \textit{dependent region (Region $4$)} that cannot function as a standalone microgrid. In this manuscript, by starting with the modified IEEE 123-node test system developed in \cite{schneider2023adaptive}, we convert some of the GFLs to GFMs to demonstrate the uniqueness of the proposed DAC. The various use cases in the case studies focus on Microgrids $2$ and $3$ working together to power Region $4$, leveraging its abundant solar energy as part of the black start assist/ Region $4$ re-energizing use case. These inverters do not actively regulate voltage or frequency, but the PID control does act to maintain maximum active power output. Furthermore, all the GFL inverters are compliant with IEEE 1547-2018 standards. 

\begin{figure}[h!]
    \centering
    \includegraphics[width = 0.95\columnwidth]{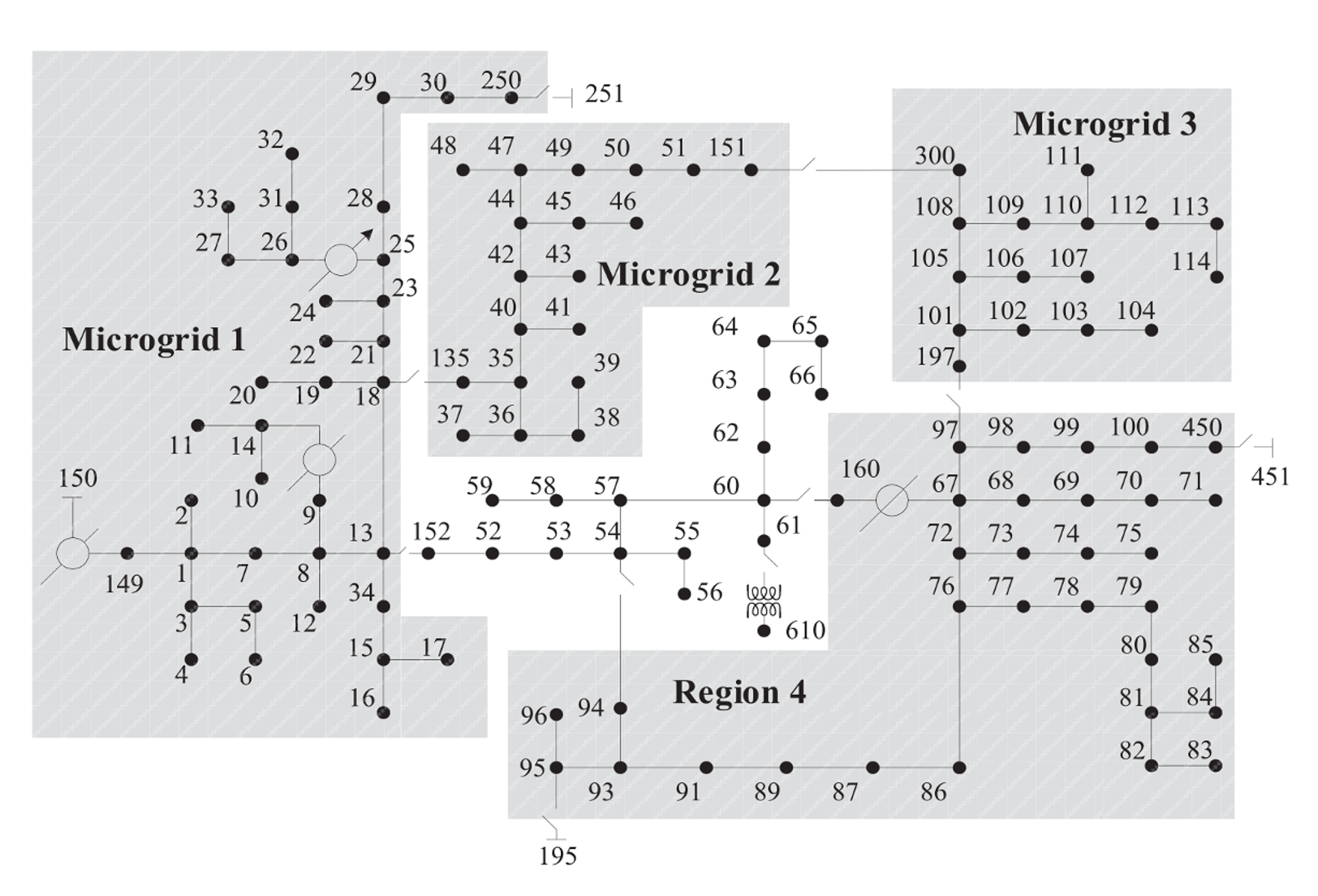}
    \caption{Illustration of the IEEE 123-node system identifying the 4 regions of the microgrids.}
    \label{fig:123bus}
\end{figure}

For the various case studies detailed below, there are small modifications done to the resources in terms of the number of  DGs, GFMs and GFLs and their corresponding capacities to demonstrate specific unique functionalities of the developed controls under some limiting scenarios. We will focus on Microgrid $2$ and $3$ resources along with resources in Region $4$.



\subsection{Performance of DAC During System Transients}
\begin{figure*}[t!]
  \centering
  \subfigure[System frequency in Microgrids $2$ \& $3$.]{\includegraphics[width = 0.31 \textwidth]{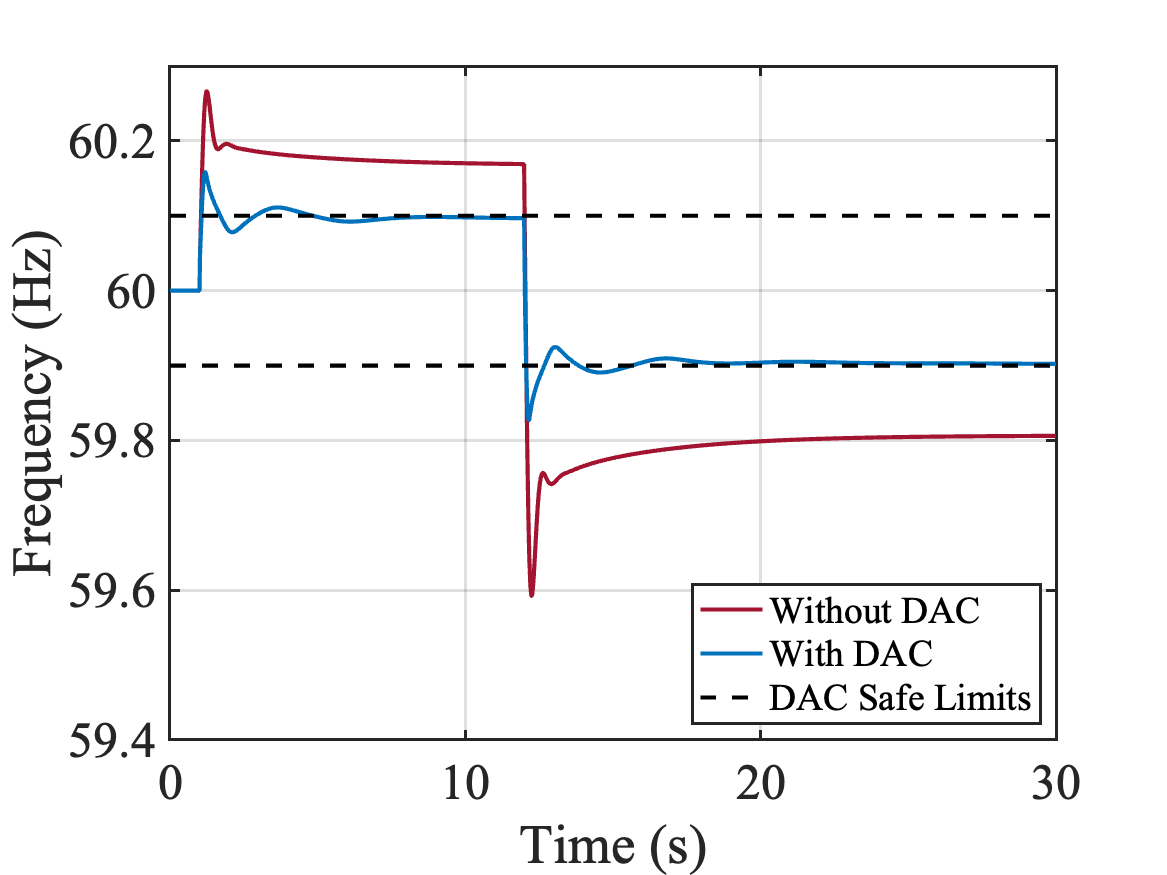}}
  \quad \subfigure[Inverter set-points \& power outputs without DAC]{\includegraphics[width = 0.31 \textwidth]{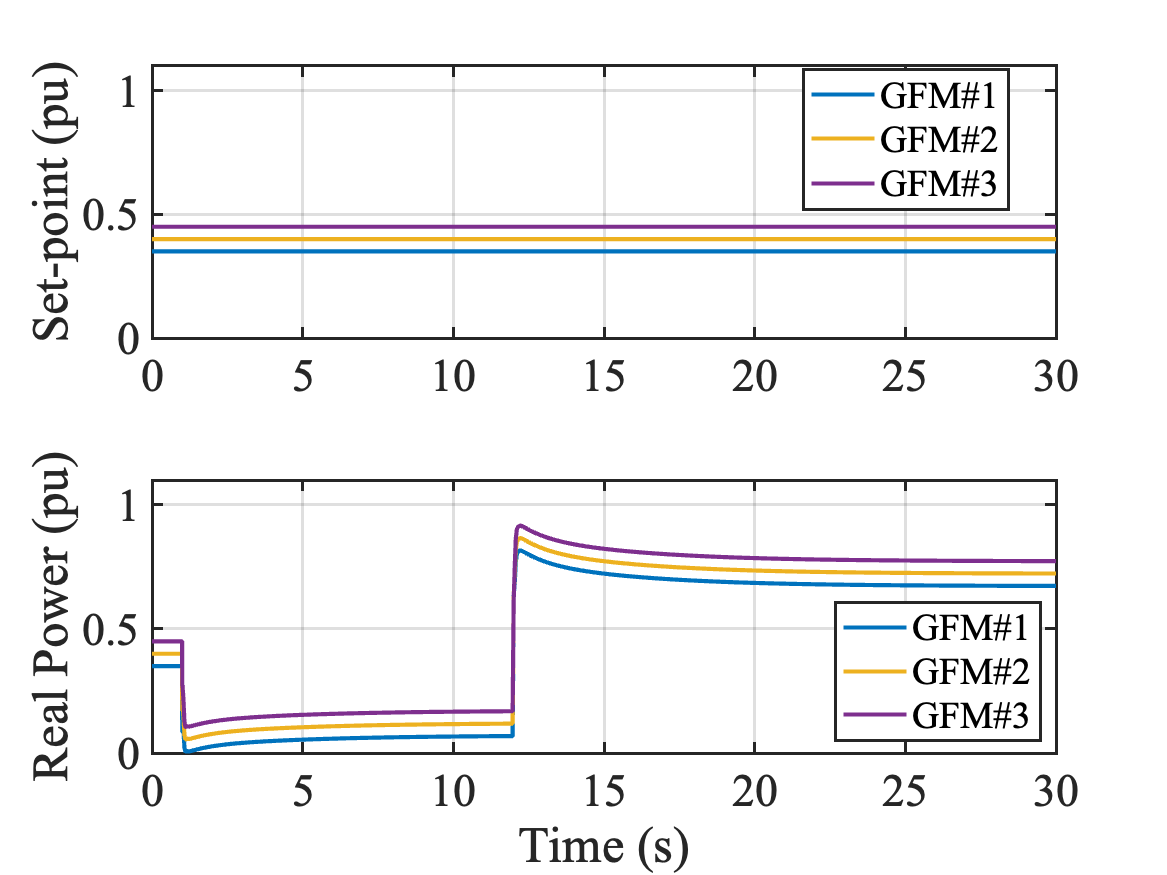}}
  \quad \subfigure[Inverter set-points \& power outputs with DAC]{\includegraphics[width = 0.31 \textwidth]{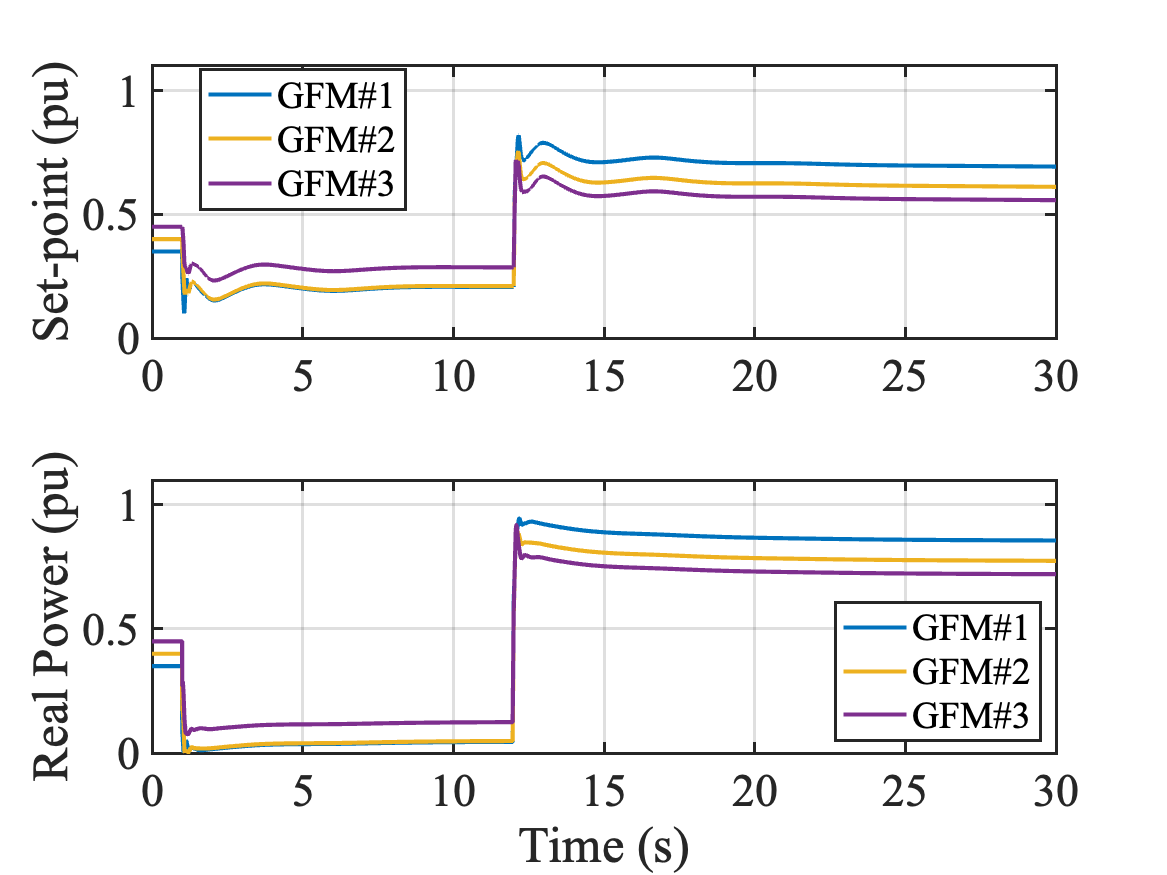}} 
  \caption{Results demonstrating the performance of DAC and its importance in maintaining the safe operation of the system.  }
  \label{fig:perf_DAC}
\end{figure*}
\begin{figure}
    \centering
    \includegraphics[width=0.95\columnwidth]{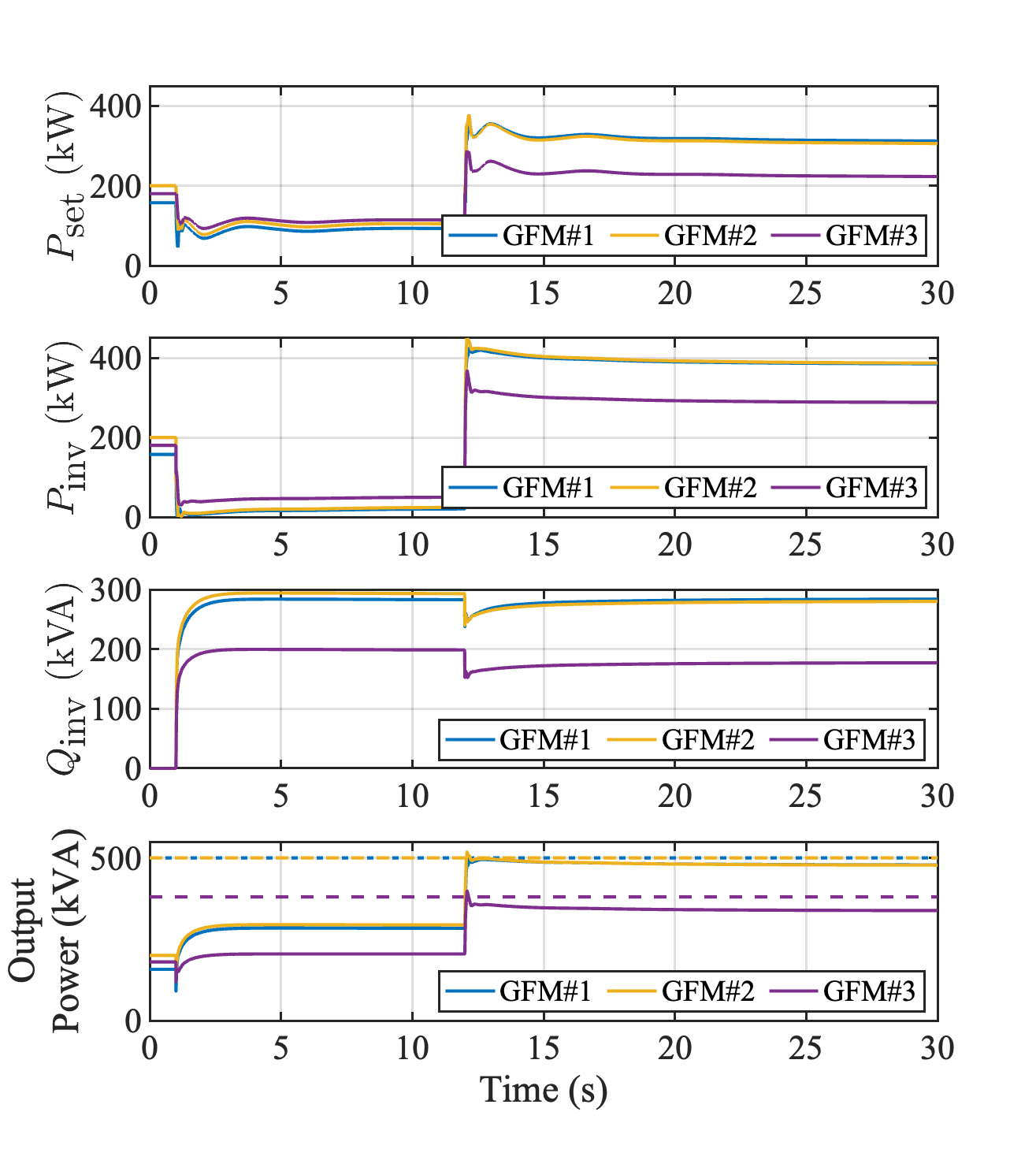}
    \caption{Real power set-point, real, reactive, and apparent powers at all the 3 GFMs. The dotted line in the bottom plot indicates the inverter size ($S_{\textrm{inv}}$).}
    \label{fig:scnE_all_powers}
\end{figure}

{To demonstrate the performance of the DAC during system transients, a simulation study was performed with the test system described above. The subsystems studied were Migrogrids 2 and 3 along with Region 4.} The resources in these regions consist of two DGs (total capacity of $2$ MVA), three GFMs (total capacity of $1.3$ MVA), and two GFLs (total capacity of $220$ kVA). Three GFM inverters were used to demonstrate the functionality of DAC using more controllable resources. 

From the simulation results, it is clear that with just the droop controls, there is no assurance that the frequency will be within the pre-defined safe limits. However, with DAC, it was observed that the safe limits are maintained ensuring local resilience. Fig.~\ref{fig:perf_DAC} shows the frequency response and inverter outputs for the dynamic event with and without DAC. The microgrids are formed at $t=1s$ such that Region $4$ is disconnected and Microgrids $2$ and $3$ operate together. The generation-load imbalance causes the frequency to increase as a result of reduced load for the generation resources. At $t=12s$, Region $4$ is connected to Microgrid $3$. At this instant, the connection of Region $4$ creates a large transient that increases the load on the generation resources, causing the frequency to dip. The GFM inverters always have their droop controls active. The safe frequency limit for this case was chosen to be $59.9$ Hz to $60.1$ Hz. 

\subsubsection{Results without DAC}
For the case without DAC, when the microgrids are formed at $t=1s$ as a result of islanding, the droop controls activate and reach a steady state with the frequency settling above the (upper) safe limit. It can be seen from Fig.~\ref{fig:perf_DAC}(b) shows that there is no change in the set-points of the GFM inverters. The outputs from the inverters are reduced due to droop controls. At $t=12s$, when the Region $4$ is connected, a large amount of load gets connected to the system (Microgrids $2$ and $3$). This results in a frequency dip and the droop controls push the GFM outputs to increase causing the frequency dip to eventually settle to about $59.8$ Hz, which is lower than the lower limit of the safe limits.

\subsubsection{Results with DAC}
In the case of activating the DACs, the droop control functionality is enhanced, which will utilize the GFM inverters to try and maintain the system frequency within the safe limits. The results in Fig.~\ref{fig:perf_DAC}(a) show that the system frequency in microgrids $2$ and $3$ are maintained within the safe limits for both the transients at $t=1s$ and $t=12s$. In both these transients, the developed DAC reacts immediately once the frequency violates the safe limits and adjusts the inverter set-points to ensure frequency resilience. Fig. \ref{fig:perf_DAC}(c) shows the inverter set-points and inverter outputs that honor the set-points to ensure the system frequency does not go beyond the safe limits. 

It is important to note that the presence of DAC arrests the frequency overshoot and the frequency NADIR is much closer to the safe limit during both events. This demonstrates that, by modifying the set-points accordingly, the DAC ensures the effective utilization of available resources to maintain local resilience. 

\subsection{Impact of DAC on System Performance under a Large Disturbance}

The DAC can prove vital when the system experiences large disturbances as they were able to guide the traditional primary (droop) controls by modifying the set-points accordingly to steer the system away from eventual instability. One such limiting scenario is considered in this subsection. 

For this scenario simulation, in the region of interest, the resource mix in the test case consists of three DGs, one GFM inverters, and three GFL inverters. The GFL inverters in the system comply with IEEE std. $1547-2018$ fault ride through (FRT) requirements and have the FRT relays set up to trip them in the must-trip zones \cite{ieee1547}. 

\begin{figure*}[h!]
  \centering
  \subfigure[System frequency in Microgrids $2$ \& $3$]{\includegraphics[width = 0.31 \textwidth]{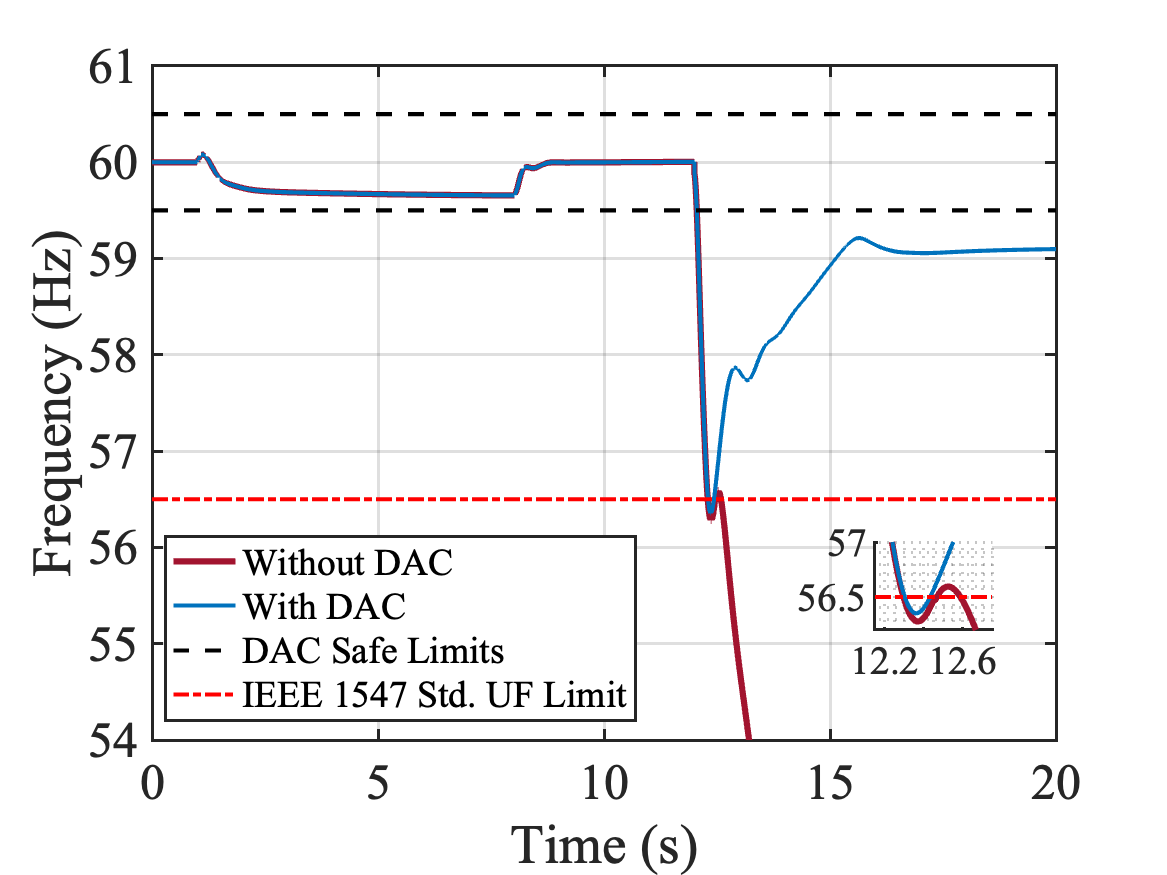}}
  \quad \subfigure[GFM inverter set-points and power output with and without DAC]{\includegraphics[width = 0.31 \textwidth]{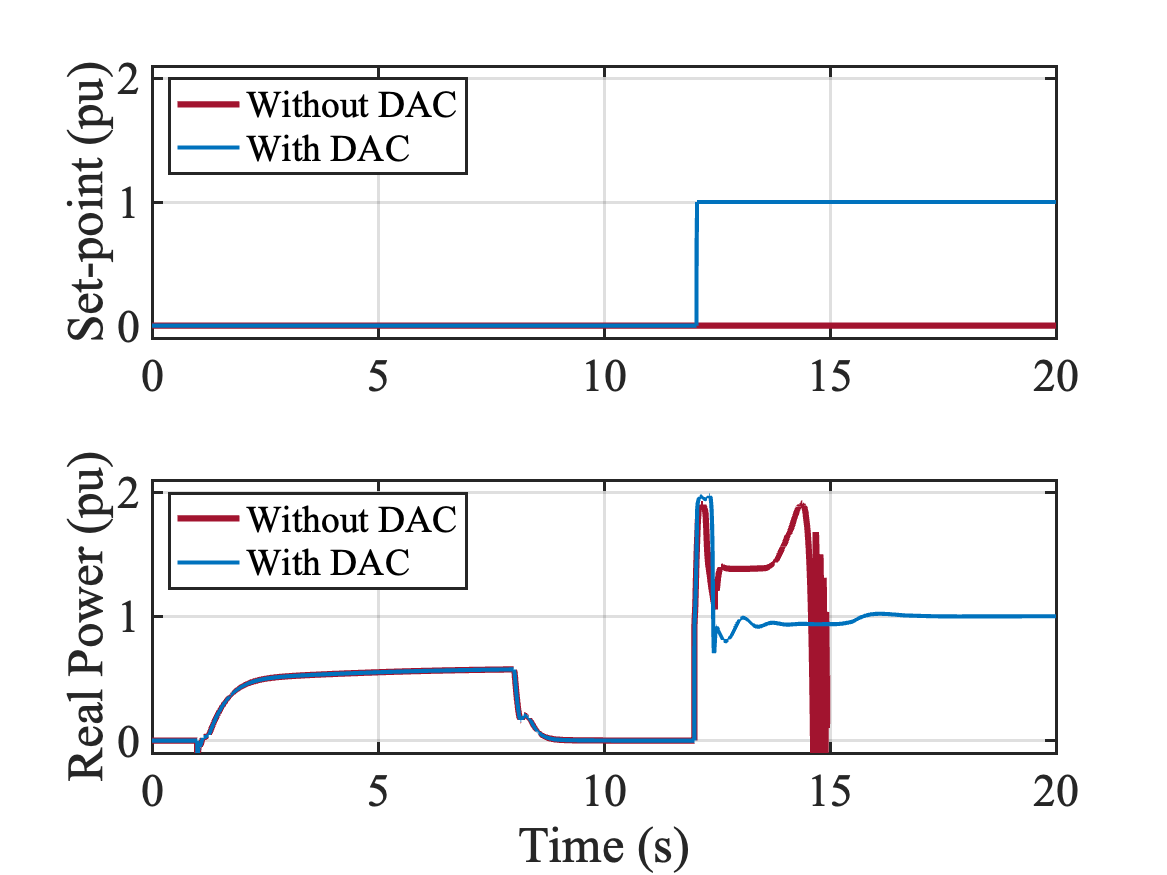}}
  \quad \subfigure[Total GFL inverter output for the cases with and without DAC]{\includegraphics[width = 0.31 \textwidth]{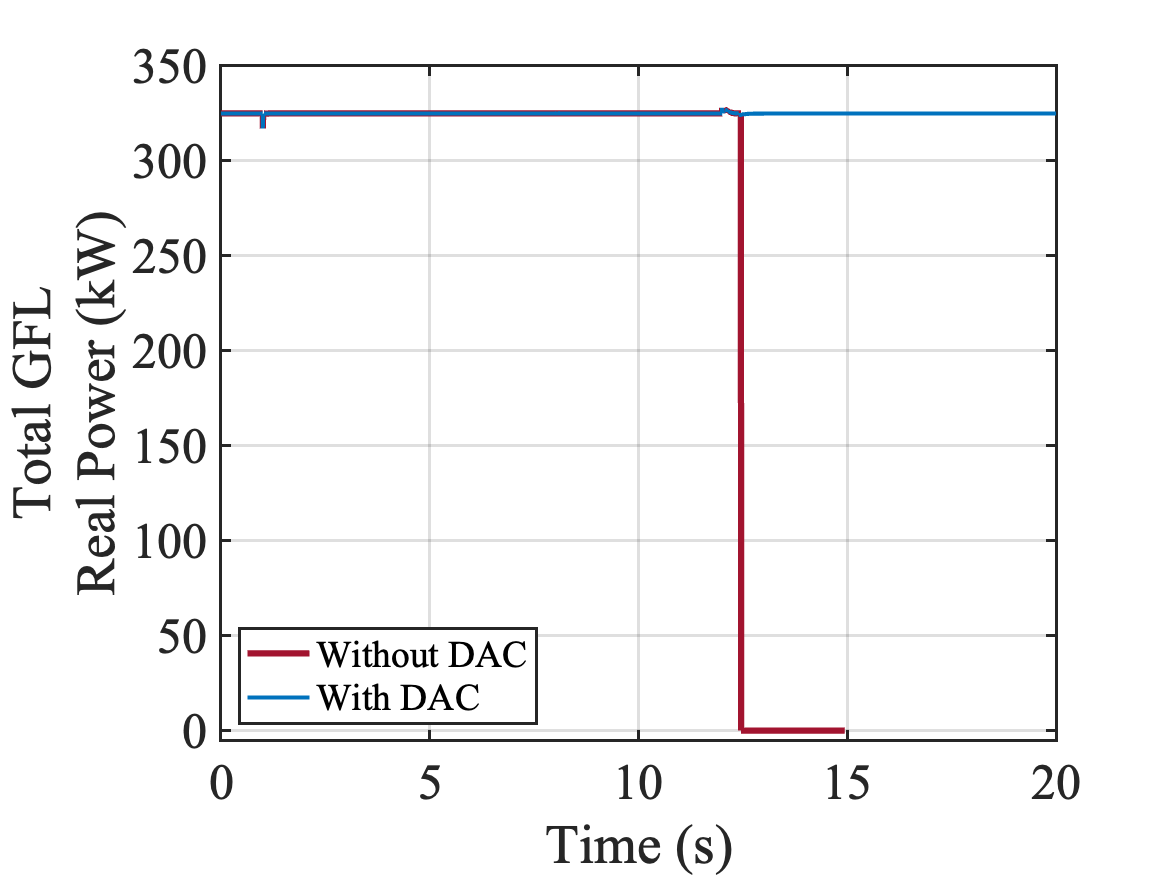}} 
  \caption{Re-energization of Region $4$ at $12s$ without DACs results in tripping of all three GFL inverters according to the IEEE 1547 standard \cite{ieee1547}. }
  \label{fig:without_and_with_controls}
\end{figure*}

Microgrids $2$ and $3$ work together to energize Region $4$, by leveraging its abundant solar energy as a part of black start assist. Both the microgrids are operating at nearly maximum solar power output, with DGs providing an additional $997$ kVA of backup. The GFM inverter in Region $3$ (GFM\#2) is modeled as a reserve resource rated $75$ kVA. The Region $4$ load is $920$ kVA, which means there is resource adequacy from a capacity perspective to energize Region $4$. However, the case study reveals additional resource utilization to handle the dynamics of the switching transient when the Region $4$ is connected.

The black start assist use case is simulated to demonstrate the performance of the proposed DACs under the described extreme switching transient. The simulation setup includes islanding the networked microgrids discussed in the previous subsection IV-A at $t=2s$ to form microgrids in Regions $2$, $3$ and $4$. At $8s$, the DGs are re-dispatched to bring the frequency of microgrids $2$ and $3$ to the nominal $60$ Hz. The GFM inverter is equipped with the traditional GFM droop controls that help respond to voltage and frequency changes in the system. The re-energization of Region $4$ is attempted at $t=12s$.



The simulations are performed with and without DAC to compare the DAC functionality for overall system performance improvement. 
Fig.\ref{fig:without_and_with_controls}(a) shows the frequency safe limits and the IEEE $1547-2018$ FRT threshold for GFL to trip if the system frequency goes below $56.5$ Hz for $160$ ms \cite{ieee1547}. Fig.\ref{fig:without_and_with_controls}(b) shows the inverter set-points and GFM inverter outputs and Fig. \ref{fig:without_and_with_controls}(c) shows the combined GFL inverter outputs.

\subsubsection{Results without DAC}
For this simulation without DAC, the frequency response of the system from Fig.~\ref{fig:without_and_with_controls}(a) shows that the frequency excursion after the Region $4$ connection at $t=12s$ lasts long enough to trip the GFL inverters after riding through the fault as per the IEEE $1547$ FRT guidelines \cite{ieee1547}. Fig.~\ref{fig:without_and_with_controls}(c) shows that the output powers of GFL inverters drop to zero as a result of tripping of the GFL inverters. The tripping of the GFL inverters results in a significant reduction in generation in the system, which eventually results in the collapse of the entire system of networked microgrids. Fig.~\ref{fig:without_and_with_controls}(b) shows how the GFM inverter struggles to keep up after the GFL inverters trip, and eventually has very large power oscillations just before the system collapses and simulation blows up.

\subsubsection{Results with DAC}
The co-simulation with DAC results in a stable system operation. The system's responses and results of co-simulation are also seen in Fig.~\ref{fig:without_and_with_controls}. When the frequency excursion goes beyond the safe region, the developed DAC attempts to bring the frequency within the safe limits (Fig.~\ref{fig:without_and_with_controls}(a)). In this case, the frequency excursion is well beyond the safe limits, which makes the DAC issue the real power set-point to the GFM inverter, $P_\textrm{set} = 1$. This overrides the inverter's default droop controls, which would make the inverter respond slower without the DAC. However, with DAC, as the GFM inverter starts responding sooner with its maximum output, the frequency NADIR is minimally improved and can be seen in Fig.~\ref{fig:without_and_with_controls}(a). This fast response of the GFM inverter due to DAC helps to improve the frequency response, and in this case, just sufficient to prevent the GFL inverters from tripping. Fig.~\ref{fig:without_and_with_controls}(c) shows that the GFL inverters do not trip in the presence of DAC.

\begin{figure}[h!]
    \centering
    \includegraphics[width = 0.95\columnwidth]{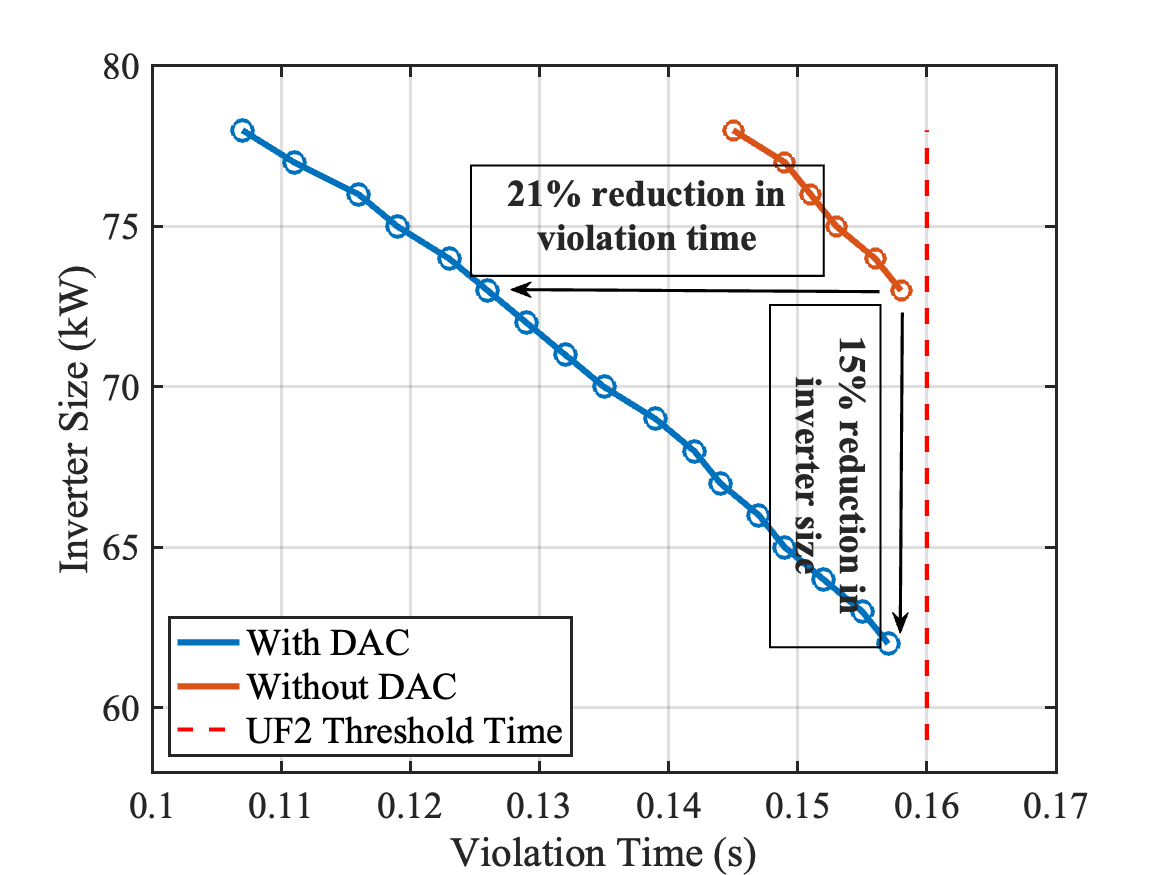}
    \caption{Impact of DACs: With DACs at inverters, one could operate with less inverter size to achieve similar performance or achieve robustness while working with the same inverter size. }
    \label{fig:Inv_size_vs_vt}
\end{figure}

In this scenario of Region $4$ re-energization, maintaining system operation with DAC is encouraging. To further understand the role of these controls and quantify the advantages of using DAC, we repeated the simulations by considering a range of inverter (GFM\#1) sizes and compared them against the case without DACs. Fig.~\ref{fig:Inv_size_vs_vt} shows how, as a function of the change in the inverter size, the violation time changes. The violation time indicates how long the frequencies were below the $56.5$ Hz threshold. Clearly, with smaller inverter sizes below, $62$ kW, even in the presence of DAC, when Region $4$ is re-energized, the combined system cannot manage the power generation and demand resulting in an unstable system. On the other side, for inverter size above $73$ kW, even the droop controls (without DAC) stabilize the system when Region $4$ is connected. {This shows that with the DAC, working with an inverter whose size is nearly 15$\%$ will maintain the same violation time.} From a different perspective, while keeping the inverter size the same, with DAC, we could achieve a $21\%$ reduction in the violation time compared to without DAC thereby enhancing the robustness of the system performance. This demonstrates again the efficient utilization of available resources by DACs to maintain stability and empower the grid under various operating conditions. 

\subsection{Cyber-Attack Resilience with DAC}
The system under consideration was modified slightly to demonstrate the cyber-attack resilience of the DAC. The regions of interest are Microgrids $2$ \& $3$ along with Region $4$. The resources in these regions consist of two DGs (total capacity of $2$ MVA), three GFMs (total capacity of $1.3$ MVA) and two GFLs (total capacity $220$ kVA) like in the case of Section \ref{sec:sim_study}-B. The number of GFM inverters is made three to demonstrate the impact of cyber-attacks on some GFM inverter power set-points and its impact on the system performance with and without DAC. The dispatch to the microgrids is controlled by a secondary control algorithm that is described in \cite{singhal2022consensus}.

\begin{figure*}[h!]
  \centering
  \subfigure[Frequency in Microgrid $2$ \& $3$]{\includegraphics[width = 0.45 \textwidth]{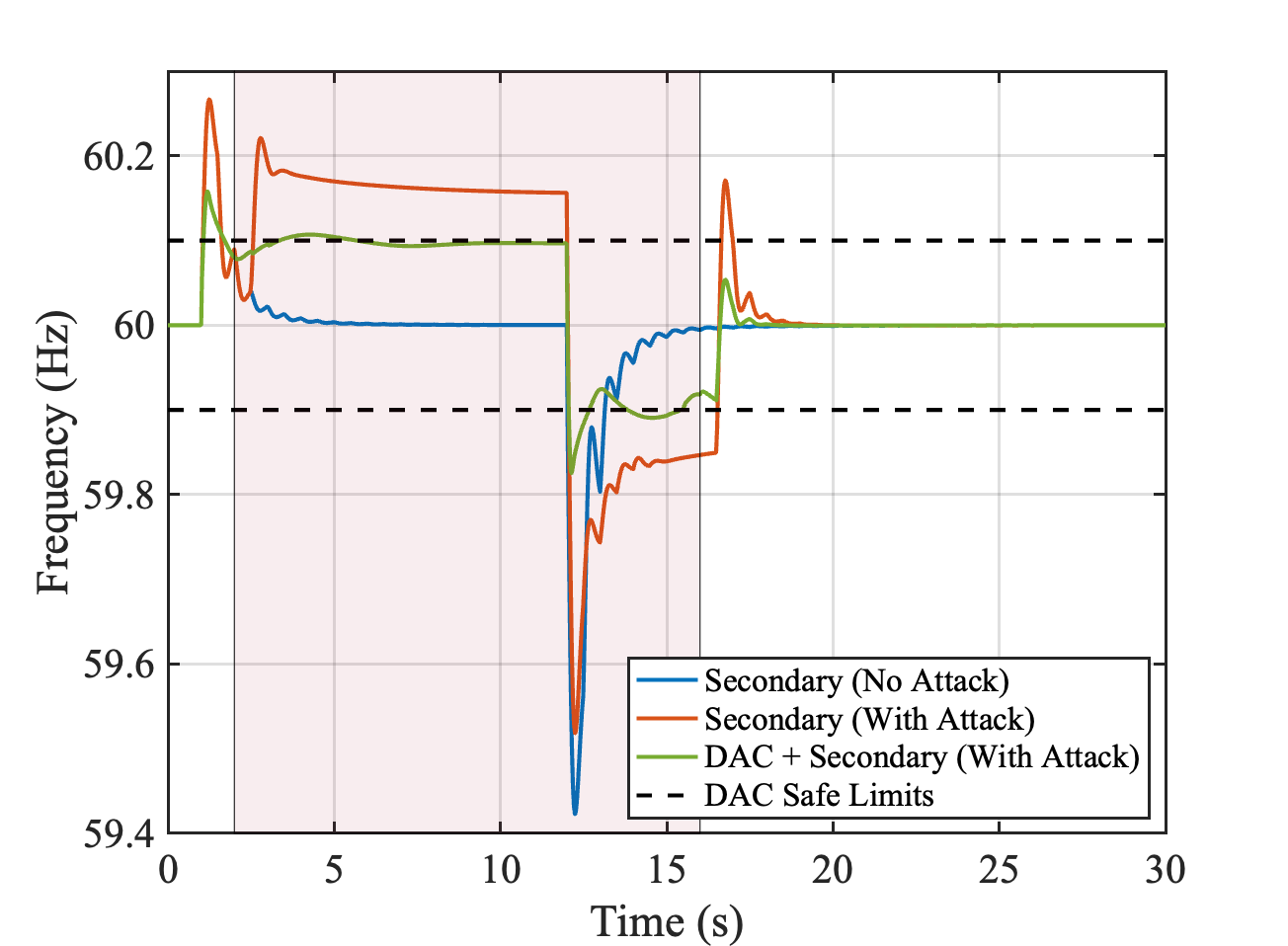}}
  \quad \subfigure[GFM inverter set-points and power outputs for the case with no cyber-attack]{\includegraphics[width = 0.45 \textwidth]{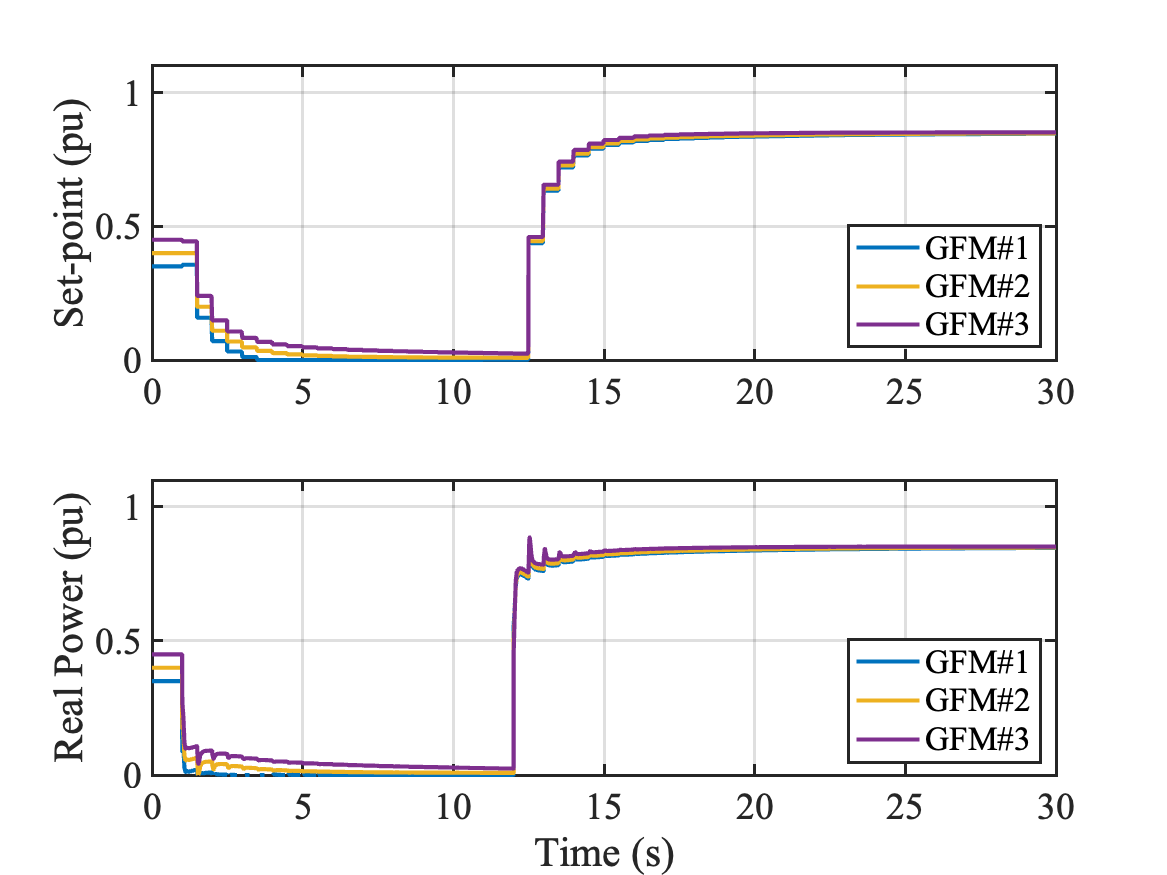}}
  \quad \subfigure[GFM inverter set-points and power outputs for case with only secondary controls during cyber-attack]{\includegraphics[width = 0.45 \textwidth]{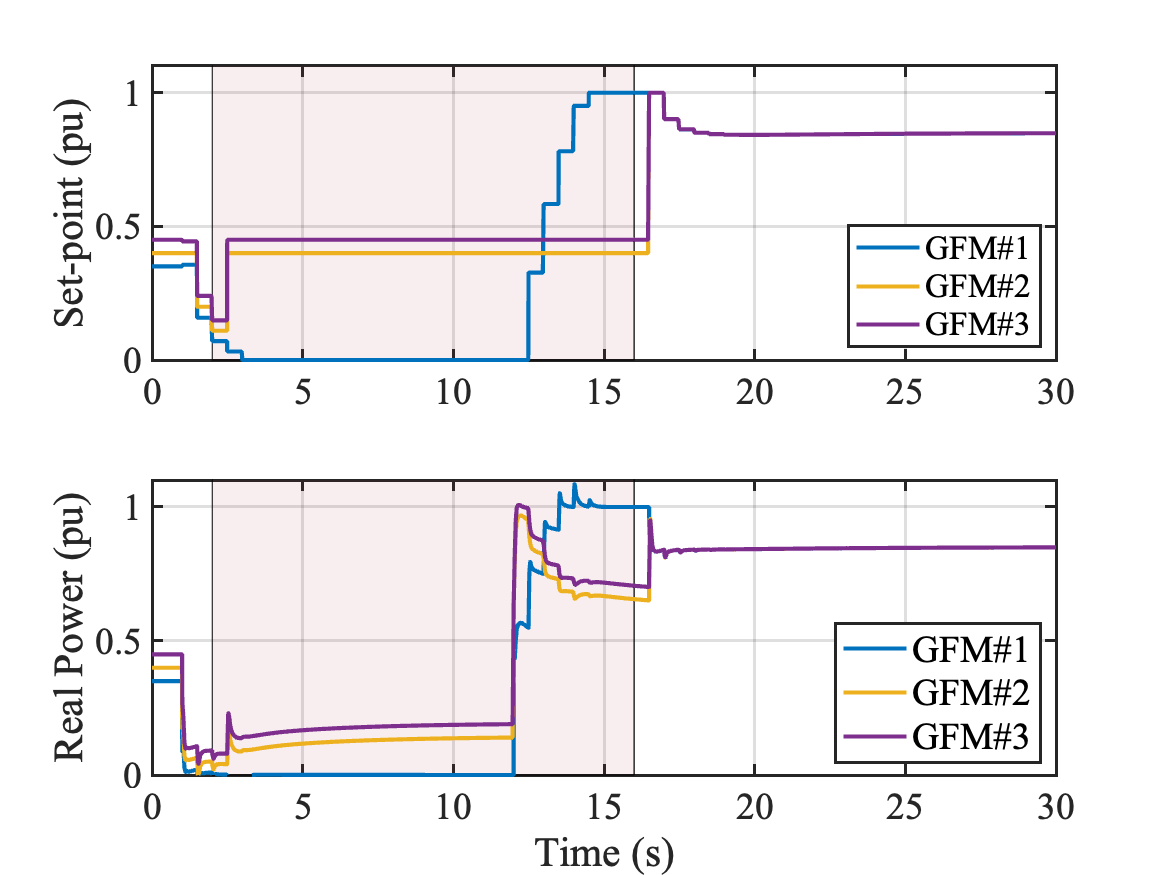}} 
  \quad \subfigure[GFM inverter set-points and power outputs for case with DAC and secondary controls during cyber-attack]{\includegraphics[width = 0.45 \textwidth]{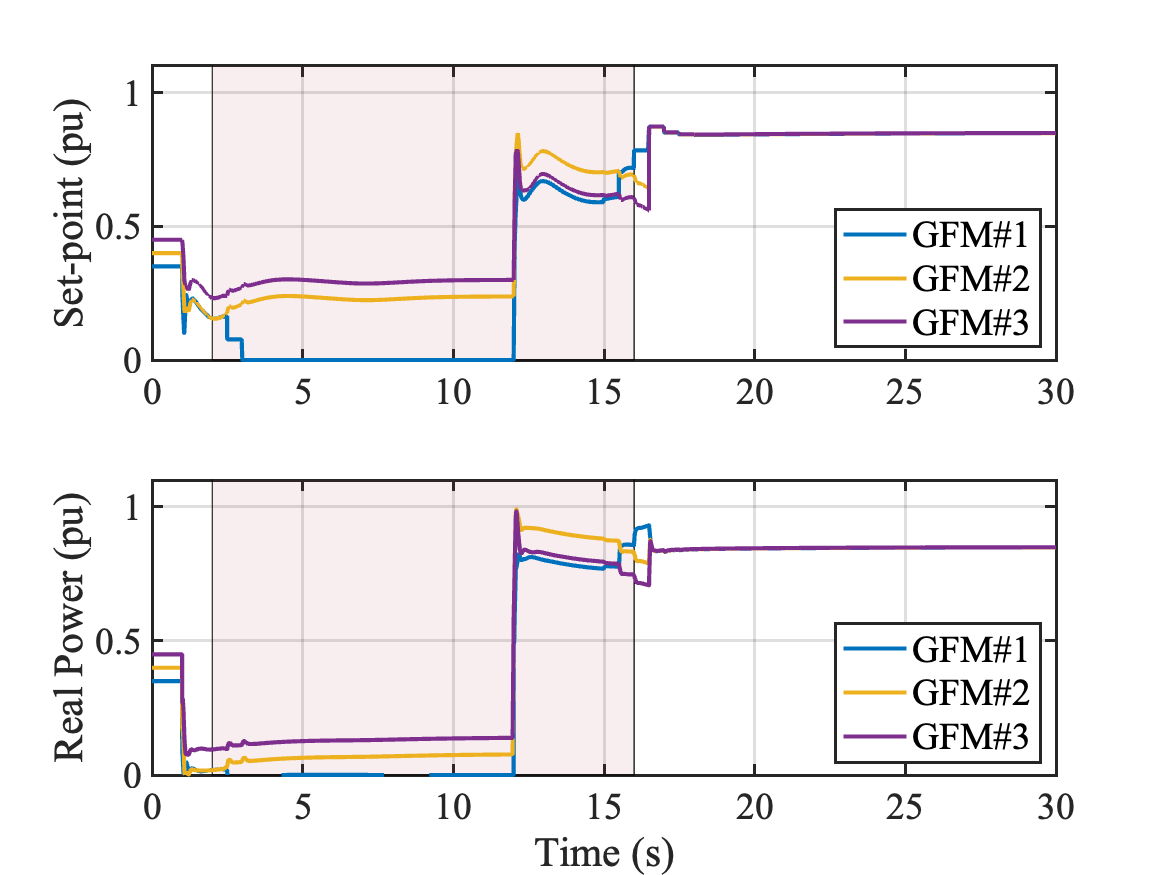}} 
  \caption{Simulation results showing the cyber-attack resilience provided by the developed DAC (cyber attack duration is from $t=2s$ to $t=16s$). The Microgrids $2$ \& $3$ are formed at $t=1s$, and Region $4$ is energized at $t=12s$.}
  \label{fig:ar}
\end{figure*}

Since the DACs use local measurements, they can prevent any malicious secondary control signals that can result in the system frequency going beyond the safe limits. Most of the GFM primary controls like droop controls cannot perform this action and will usually directly honor the secondary control dispatch. Fig.~\ref{fig:ar} shows the results of the two cases with and without DAC that demonstrate the attack resilience that comes with the proposed DAC.

The nature of the cyber-attack is described as follows. The pre-disturbance set-points are treated by the attacker as nominal set-points and during the attack window, the attacker simply masks the actual secondary control setpoints provided to the inverters and instead sends the nominal set-points irrespective of the underlying disturbance on the system. As a result of this, depending on the type of disturbance the system is undergoing, the impact on the system will be different.

In Fig.~\ref{fig:ar}, the results show that in the absence of DAC, during the cyber-attack from $t=2s$ to $t=16s$, the GFM inverter set-points are impacted. The case study is focused on bringing out the capability of the DAC due to the use of local measurements but is not focused on the details and modality of the cyber-attack. The inverter set-point modification emulates a cyber-attack to change the power outputs of the GFM inverters. In this scenario, GFM\#2 and GFM\#3 are attacked and GFM\#1 inverter set-points are not attacked. This allows us to visualize how the system components respond individually in the presence and absence of cyber-attacks. All the cases described below have GFM droop controls active. 

The ``Secondary (No Attack)'' response in Fig. \ref{fig:ar}(a) and Fig. \ref{fig:ar}(b) show the results for a case of deployment of the secondary controls under no cyber-attack scenario. This case is that of an ideal operating condition where the GFM droop controls with the secondary controls from \cite{singhal2022consensus} can maneuver the system to achieve a stable $60$ Hz (nominal frequency) operation. However, these controls cannot function as expected during an attack on them, and as a result, they are not capable of ensuring that the frequency goes back to nominal or maintains the safe region. The next subsections discuss the results with and without DAC to demonstrate the versatility of the developed DAC that can alleviate the impact of a cyber-attack.

\subsubsection{Results without DAC}
The results show that in the absence of DAC, the GFM inverter outputs vary as per the secondary control set-points irrespective of the safe limits. It is important to note that the inverters have the default droop controls as their primary controls are active throughout the simulation. As the inverter responds to set-point increase (due to the cyber-attack) as shown in Fig.~\ref{fig:ar}(c), the frequency in the system increases to go beyond the safe limit during $t=2s$ to $t=12s$ (Secondary (With Attack) case in Fig. \ref{fig:ar}(a)).  The key functionality of the secondary controls is to maintain the nominal frequency of $60$ Hz, however, due to the compromised set-points, the frequency is not $60$ Hz and goes beyond the safe limits as well. It is also interesting to see that the GFM\#1 inverter's output powers go to $P_\textrm{set}=0pu$ trying to bring down the frequency between $t=2s$ to $t=12s$. The GFM inverter cannot lower its power output any further in this operating scenario.

At $t=12s$, the system undergoes the Region $4$ re-energizing transient, where the frequency drops, but the GFM\#2 and GFM\#3 inverters do not change the power outputs as the set-points were under attack until $t=16s$, and with the incorrect/inappropriate set-points from the secondary controls, these inverters fail to modulate the power output in response to the transient. This results in a frequency dip even though the GFM\#1 secondary control set-point goes to $P_\textrm{set}=1pu$. However, after the attack duration, the secondary controls operate normally to increase the GFM\#2 and GFM\#3 inverter outputs to bring the system frequency back to $60$ Hz. It is important to observe that in the absence of DAC, during the attack duration, the system was operating beyond the safe region.

\subsubsection{Results with DAC}
In the case where the DAC is active along with the GFM droop during the cyber-attack, the DAC senses that the frequency goes beyond the safe limits during both events due to the set-point attack. The DAC then appropriately modifies the set-points to maintain the system frequency within the safe limits. In the case with DAC, Fig.~\ref{fig:ar}(d) shows that the DAC acts even during the cyber-attack on the secondary control set-points that are compromised. In this case, the DAC adjusts the set-points ensuring that the frequency excursion is not beyond the safe limits. The purpose of the DAC is to ensure that the system operates within the safe operating region and therefore adjusts the set-points of the GFM inverters to just bring the system frequency inside the safe region, thereby demonstrating local resilience. 

During normal operations, it is not obvious to include the DAC as there are no indications of unsafe operations; however, during an abnormal condition like a cyber-attack as described, the DAC's capability shines to provide attack resilience with the decentralized and autonomous nature of the DAC. 

The different models used in the validation study and the observations made are included in Table~\ref{table_summary}.

\begin{table*}[ht!]
\caption{Description of the different use cases utilized in the validation study. }
\label{table_summary}
\begin{center}
\begin{tabular}{|M{3 cm}|M{2 cm}|M{5 cm}|M{5 cm}|}
\hline
\textbf{Case} &\textbf{ \# INVs, \# DGs} & \textbf{Disturbance} & \textbf{Observation} \\ \hline
\textbf{Subsection IV-B}  Control performance during transients &
  2 DGs, 3 GFMs and 2 GFLs &
  The system is islanded at t = 1s and the region 4 is disconnected. The microgrid 2 and 3 form a networked microgrid. Again, at t = 12s, region 4 is energized by the networked microgrid. &
  Frequency limits were maintained using DACs whereas it violates without DAC \\ \hline
\textbf{Subsection IV-C} IEEE 1547 Compliance at GFL Inverters &
  3 DGs, 1 GFM and 3 GFLs &
  The system is islanded at t = 2s and the region 4 is disconnected. The Microgrid 2 and 3 form a networked microgrid. At 8s, the DGs are re-dispatched to bring the frequency of microgrids 2 and 3 to the nominal 60Hz. At t = 12s, region 4 is energized by the networked microgrid. &
  The DACs maintain the UF2 threshold \textbf{avoiding GFL tripping} whereas without DAC, all 3 GFLs trip. \\ \hline
\textbf{Subsection IV-D} Cyber Attack on set-points dispatched by centralized control &
  2 DGs, 3 GFMs and 2 GFLs &
  Besides the disturbance created in Subsection IV-B, between t = 2s and t = 16s, there is an attack on the secondary control inverter set-points. &
  The DACs modify the attacked set-points to \textbf{achieve resiliency} whereas without DAC the attack results in larger frequency deviations and unintended control effort \\ \hline
\end{tabular}
\end{center}
\end{table*}

\section{Discussion and Conclusions}
\label{sec:Con}
Over the recent decades, there has been a rapid evolution of technologies and increased deployment of power electronics-based inverters, especially GFM inverters. While many system-level controls aim at enhancing performance under normal operating conditions, this paper presents safety-based DACs to enhance the system performance under normal and stressed conditions. The DAC operates when the system frequency goes beyond the safe limits. {The DACs reside at the device level (at the GFM inverters) and come in between the secondary and primary controls in the hierarchy of controls. They only act by modifying the secondary control set points to maintain the local frequency constraints, and otherwise just pass the secondary control set points to the inverters.} The DACs work with the droop controls in a complementary manner to enhance the droop controller functionality during emergencies. The DAC is demonstrated as a safety-promoting resilience control and the safe limits can be adjusted along with the control parameters to enhance the effective utilization of the GFM resources. 

The DAC is an effective way to maximize resource utilization that ensures the system does not go out of the safe region and in limiting cases can prevent a collapse of the system due to cascaded contingencies. In this paper, a scenario is presented where DACs prevent IEEE $1547$ compliant GFL inverters tripping during a large disturbance \cite{ieee1547}. The key impact of the DAC is demonstrated with the maintenance of the system frequency within the safe limits when resources exist and help drive the system toward the safe limits with the needed urgency when there are large disturbances. The DAC relies on the resource nameplate details and local measurements to steer the GFM resources to enhance the full utilization of system resources. This enables the DAC to have a check in place that can override the secondary control if the system steers away from the safe region. Overall, the DAC enhances the GFM inverter capabilities and improves its droop functionality. The proposed controls are applicable universally and the team will implement these controls by including other resources like GFL inverters. This work will also be explored for applications beyond microgrids in future studies. 



\end{document}